\documentclass{article}

\PassOptionsToPackage{numbers, compress}{natbib}

\usepackage[preprint]{neurips_2024}

\usepackage[utf8]{inputenc} 
\usepackage[T1]{fontenc}    
\usepackage{hyperref}       
\usepackage{url}            
\usepackage{booktabs}       
\usepackage{amsfonts}       
\usepackage{nicefrac}       
\usepackage{microtype}      
\usepackage{xcolor}         
\usepackage{graphicx}
\usepackage[numbers]{natbib}

\title{Turkey’s Earthquakes: Damage Prediction and Feature Significance Using A Multivariate Analysis}

\author{%
   Shrey Shah\\
   \texttt{shreyshah1011@gmail.com} \\
   \And
   Alex Lin  \\
  \texttt{emailtoalex14@gmail.com} \\
   \AND
   Scott Lin \\
   Foothill High School \\
   Pleasanton, CA \\
   \texttt{scottlinca@gmail.com} \\
   \And
   Josh Patel \\
   The Peddie School\\
   Hightstown, NJ \\
   \texttt{joshbpatel@gmail.com} \\
}

\author{Shrey Shah\thanks{Lead Authors}  \hspace{1cm} Alex Lin\footnotemark[1] \hspace{1cm} Scott Lin \hspace{1cm} Josh Patel \\
{\bf Michael
Lam} \hspace{1cm} {\bf Kevin Zhu} \\
        Algoverse AI Research \\
        \texttt{shreyshah1011@gmail.com, kevin@algoverse.us}}

\begin{document}

\maketitle

\begin{abstract}
  Accurate damage prediction is crucial for disaster preparedness and response strategies, particularly given the frequent earthquakes in Turkey. Utilizing datasets on earthquake data, infrastructural quality metrics, and contemporary socioeconomic factors, we tested various machine-learning architectures to forecast death tolls and fatalities per affected population. Our findings indicate that the Random Forest model provides the most reliable predictions. The model highlights earthquake magnitude and building stability as the primary determinants of damage. This research contributes to the reduction of fatalities in future seismic events in Turkey.
\end{abstract}

\section{Introduction}

Earthquakes serve as one of the most catastrophic natural disasters, claiming over 61,000 lives in 2023 alone. \cite{ngdcwds, herece2023}Turkey, in particular, is struck by thousands of earthquakes annually because of the multiple major seismic fault lines running through its geography. Most notably, the recent 2023 Turkey–Syria earthquakes, of magnitudes 7.8 and 7.7, killed a confirmed 53,537 people.\cite{Barrons} Recent predictive models, such as the RECAST model, excel at forecasting future earthquakes, along with information such as power level, location, and time of formation. \cite{dascher2023using} However, in order to better prepare earthquake mitigation strategies, it is necessary to predict the damage outcome and feature significance of these earthquakes. Our study bridges the gap in understanding earthquake damage by predicting its severity based on major influencing factors such as building stability, earthquake depth, and population density per province. We also seek to determine the significance of these factors in contributing to the overall destruction caused by earthquakes. The primary objective of this research is to formulate a model that accurately predicts the death toll and fatalities per affected population in Turkish earthquakes by testing various machine learning architectures and using the most accurate one to assess the significance of each factor, thereby enhancing earthquake preparedness and resistance (see results).





\section{Methodology}

\subsection{Data}
In our study, we analyzed four datasets \cite{kandilli, ergin1967catalogue, usgs, https://doi.org/10.17603/ds2-8710-ad45} of earthquakes occurring prior to 1950, each containing information on magnitude, earthquake depth(km), MMI (Modified Mercalli Intensity) or shaking intensity of earthquakes, death toll, and epicenter coordinates. Additionally, we integrated a 2022 dataset on factors for each province in Turkey \cite{MMI, popdens} by matching epicenter coordinates with provinces. The factors used are population density, structural integrity and stability based on age, design, etc. of building (BCI), and the susceptibility of buildings to earthquake damage (SVI). \cite{bcisvi} We selected these variables since variations in them were hypothesized to influence death tolls. Our study requires two sets of machine learning models: one using death toll as the output variable and the other using death per capita of the affected population. Total deaths are useful in predicting the number of deaths from an upcoming earthquake in a specific region to help in immediate disaster response and resource allocation. Deaths per capita can be used for comparing the effectiveness of building codes and other infrastructural features more directly.

To calculate the death per capita of each earthquake, we first utilized Joyner and Boore's (1981)\cite{joyner1981peak} attenuation relations that predict PGA (peak ground acceleration) using magnitude and the distance from the fault rupture, as shown in Formula ~\ref{PGA prediction}.\cite{joyner1981peak}
\begin{equation}\label{PGA prediction}
    \log(PGA(gravity)) = b_1 + b_2M + b_3 \log(R + b_4)
\end{equation}
The values of the coefficients, \(b_1, b_2, b_3, \) and \(b_4\) in the above formula were derived from a study by Kalkan and Gülkan (2004)\cite{kalkan2004site} by performing regression analyses on ground motion data specific to Turkish seismic events. In our study, we consider individuals in an area with an MMI (Modified Mercalli Intensity)\cite{MMI} of at least IV, where earthquakes are widely felt and fatalities are more likely, as affected by the event. This threshold was chosen because fatalities are unlikely below an MMI of IV, which corresponds to a Peak Ground Acceleration (PGA) of 2.8g \cite{wald1999relationships}. By substituting the known values into the relations, the radius was found. In our study, we assume a circular pattern for earthquake damage prediction, which simplifies the complex nature of seismic wave propagation and distribution. Seismic waves radiate outward in all directions from the earthquake's epicenter \cite{usgsscience}. While this circular approximation might not capture all the intricacies of fault geometries, the impact on our model's accuracy is minimal due to the general nature of seismic wave dispersion. Using the radius derived from the attenuation relations, we applied the formula for the area of a circle to estimate the surface area of the affected region, which was then multiplied by the population density of the province where the earthquake occurred to find the total affected population. We then divided the total number of fatalities by the total affected population to calculate the death per capita for the earthquake.

\subsection{Evaluation Metrics}

In training our models, we deviate from the standard Mean Squared Error (MSE) loss function due to its interpretative limitations with our data. Instead, we utilize Mean Absolute Percent Error (MAPE), which presents loss as a percentage error rather than a numerical discrepancy, making it more suitable for our context. Additionally, we employ Mean Absolute Error (MAE) as a complementary metric. MAE is beneficial as it uniformly weights all values, irrespective of direction, providing a robust assessment of model performance.

\subsection{Model Selections}

In this study, our various machine learning models use the factors described in 2.1 as x-values and either the number of fatalities per population affected or the death toll as y-values. 

The linear regression model is this study's baseline model. It implements Ordinary Least Squares (OLS) regression to fit the dataset and generate a model summary with the metrics described in 2.2.

The neural network's architecture features an input layer, multiple hidden layers, and an output layer. The model is compiled with the Adam optimizer and MAPE loss function. The data undergoes an 80/20 train-test split, and the model is trained for 4500 epochs.

For the Decision Tree, Random Forest, Ridge, and LASSO machine learning models, the data is first split in an 80/20 train test split. Hyperparameter optimization is conducted on the training data using grid search with 5-fold cross-validation, employing negative MAPE as the scoring metric. The model is then trained and evaluated on the test data.

In the Decision Tree Regressor, the hyperparameters are the maximum depth, the minimum number of samples required to split an internal node, the minimum number of samples required to be at a leaf node, and the number of features to consider when looking for the best split. In the Random Forest Regressor, the hyperparameters are the number of trees in the forest, the maximum depth of each tree, the minimum number of samples required to split an internal node, the minimum number of samples required to be at a leaf node, and whether bootstrap samples are used when building trees. Both Ridge and LASSO Regression models use one hyperparameter for 'alpha.'


\section{Results}

The linear regression models exhibited a suboptimal performance, with a MAPE of 23.25 and a MAE of 0.0030 for per capita predictions and a MAPE of 88.92 and an MAE of 2737.01 for the total death toll. Despite the seemingly favorable per capita predictions, fitting the model directly to Ordinary Least Squares (OLS) without splitting the data into training and testing sets (as described in section 2.3) suggests potential overfitting. Table~\ref{table 1} and Table~\ref{table 2} present the testing values for differing machine learning models grouped together due to their shared evaluation methods, as detailed in section 2.3.

\begin{table}[ht!]
\centering
  \caption{Deaths per Population Predictor Evaluation Metrics}
  \label{table 1}
  \begin{tabular}{llllll}
    \toprule
      & Neural Network & Decision Tree & Random Forest & Ridge & LASSO \\
    \midrule
    Training MAPE & 85.70 & 5.79 & 7.79 & 19.39 & 35.01\\
    Training MAE  & 0.0024 & 0.0026 & 0.0026 & 0.0030 & 0.0031\\
    Testing MAPE & 83.97 & 9.92 & 10.61 & 32.09 & 31.64\\
    Testing MAE  & 0.0013 & 0.0026 & 0.0026 & 0.0022 & 0.0018\\
    \bottomrule
    \centering
  \end{tabular}
\end{table}
\begin{table}[ht!]
\centering
  \caption{Death Toll Predictor Evaluation Metrics}
  \label{table 2}
  \begin{tabular}{llllll}
    \toprule
      & Neural Network & Decision Tree & Random Forest & Ridge & LASSO \\
    \midrule
    Training MAPE & 84.37 & 3.99 & 1.76 & 35.01 & 86.29\\
    Training MAE  & 2379.86 & 464.98 & 187.83 & 2193.49 & 2453.03\\
    Testing MAPE & 83.78 & 12.16 & 6.20 & 84.91 & 63.93\\
    Testing MAE  & 2091.89 & 802.43 & 1175.13 & 3041.39 & 2858.44\\
    \bottomrule
    \centering
  \end{tabular}
\end{table}

Based on the performance metrics, the Decision Tree model and Random Forest model provide the best results due to having low MAPE and MAE values compared to the other data. The Random Forest model would be a better predictive model, not only due to the better MAPE scores, is an ensemble model combining multiple decision trees. Since the forest model averages many tree models, this leads to a lower risk of overfitting and better feature variability. Therefore, the Random Forest model is the best damage-predictive model for this study.

\begin{figure}[htbp]
    \centering
    \begin{minipage}[b]{0.48\textwidth}
        \centering
        \includegraphics[width=\textwidth]{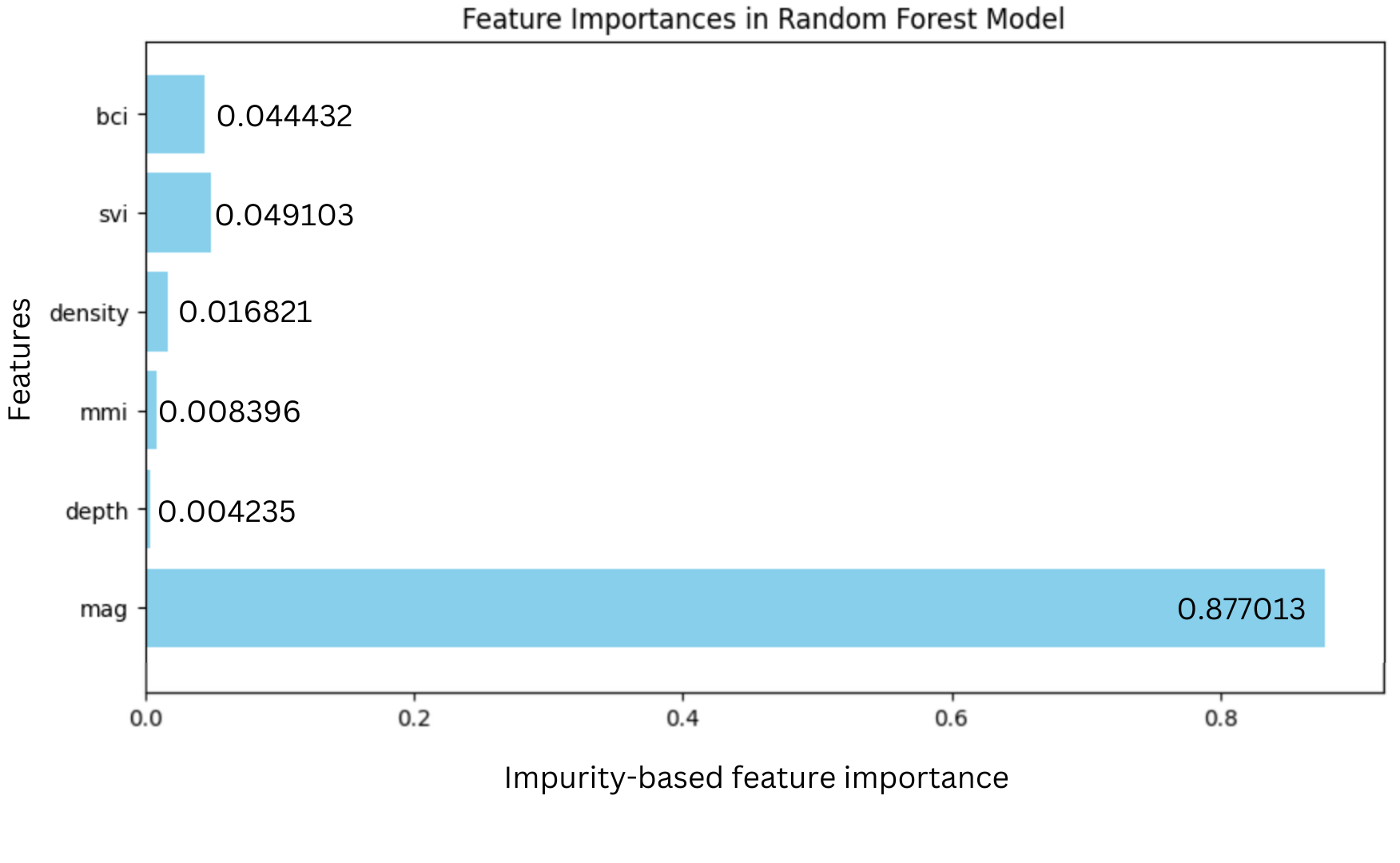}
        \caption{Factor significance in Influencing Earthquake Deaths}
        \label{fig:image1}
    \end{minipage}
    \hfill
    \begin{minipage}[b]{0.48\textwidth}
        \centering
        \includegraphics[width=\textwidth]{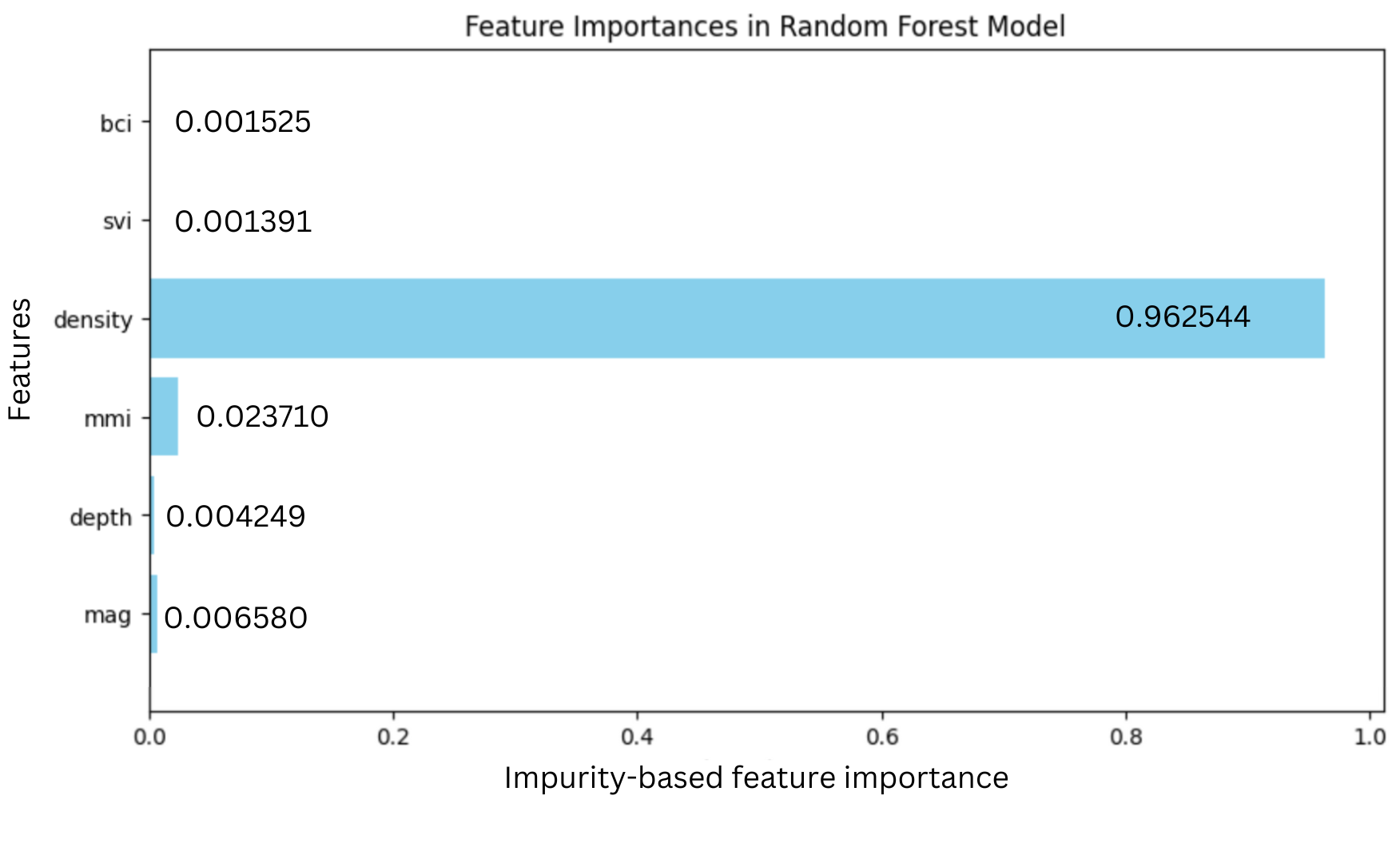}
        \caption{Factor significance in Influencing Earthquake Deaths Per Capita}
        \label{fig:image2}
    \end{minipage}
\end{figure}
In Random Forest models, factor importance applies a score to each x-coordinate, reflecting how relatively significant each factor is in predicting the damage. This is calculated by adding the reduction in error that each feature contributes at each split across all nodes and trees in the forest and then averaging these scores. The importance values are normalized to sum to one, providing a clear comparison of how much each feature contributes to the model. The feature significance of both predicting deaths and predicting deaths per population affected is shown in the graphs above.

For the model predicting deaths per population affected, the feature importance values indicate that population density is overwhelmingly the most significant factor. This is because it is directly involved in calculating the target variable, as it is multiplied by surface area affected to get the number of people affected.

In the death toll prediction model, magnitude is the most significant factor in earthquake damage. Notably, SVI and BCI have similar importance scores, suggesting that earthquake vulnerability and building stability have comparable influence on predicting the death toll. Understanding these importance scores helps in identifying which factors are most critical to focus on when developing models and strategies for earthquake impact mitigation and response planning.

\section{Discussion}
In this paper, we designed models in the field of earthquake damage prediction and determined the significance of socioeconomic factors to earthquake damage. In the future, combining the results from earthquake prediction models such as the RECAST and ETAS model could make our model extremely useful to Turkish disaster preparation agencies, as we would be predicting vital information on earthquakes to come, leading to better mitigation strategies. \cite{dascher2023using, lombardi2015estimation}

Nonetheless, there are important limitations we hope to address in future iterations of this model. First, conceptually, there are many reasons earthquake damage prediction is difficult. Often, the aftermath of the earthquake can have a more devastating effect than the seismic event itself. Unpredictable variables such as possible fires or explosions can create a difficult environment for damage prediction as it is impossible to know what other catastrophes a simple earthquake could lead to. Our model provides a useful conceptual foothold to build upon. Theoretically, as technology advances, more sophisticated methods will enable us to conquer this issue. Additionally, since we only used data preceding 1950, there were certain limitations, as most data collected on earthquakes before that point would be inaccurate or incomplete. This led to a limited training dataset of only 99 values, but all with complete data in the most important variables.

\section{Related Works}

Recent advancements in machine learning have improved the accuracy of many earthquake prediction models. These models, such as the recent RECAST models, rely on dense seismic networks and automated data processing techniques.\cite{dascher2023using} The RECAST Model utilizes deep learning neural networks to expand upon the ETAS model’s temporal point processing model. By adding features such as predicted location as well as other geophysical data, the RECAST model improves the accuracy and quantity of data in earthquake prediction. The development in machine learning and earthquake prediction technology has been paralleled by a rise in earthquake damage prediction. For example, DrivenData began hosting a damage prediction competition, where participants predict the level of damage caused to buildings from Nepal's Gorkha earthquake, based on building data and socioeconomic statistics.\cite{drivendata} However, the competition overlooks dynamic variables having the largest effects on the loss of human life. Our team realized that with different data, we could use a similar approach to predict the damage of earthquakes that will happen in the future, thus connecting earthquake prediction to damage prediction.

\section{Conclusion}
This study has concentrated on predicting earthquake damage severity in Turkey by analyzing factors such as building stability, earthquake depth, and population density. Through the evaluation of various machine learning architectures, our objective has been to accurately forecast death tolls and fatalities per affected population. Additionally, performing feature importance on the final model improves earthquake preparedness and resilience strategies. We hope that this research will prompt increased efforts toward mitigating fatalities caused by earthquakes.

\section{Acknowledgements}
All source code and the text of this paper were authored by Shrey Shah, Alex Lin, Scott Lin, and Josh Patel, who designed the project following an extensive literature review. Scott Lin and Josh Patel are high school students at Foothill High School and The Peddie School, respectively.

We extend our gratitude to Mike Lam and Kevin Zhu for their contributions through lectures on machine learning and research skills, suggested readings, high-level guidance, and constructive comments on the manuscript.

\bibliographystyle{unsrt}
\bibliography{references}

\end{document}